\documentclass[12pt,a4paper,final]{iopart}

\usepackage{iopams}
\usepackage{graphicx}
\usepackage{color}

\begin{document}

\title{Semiconductor to metal transition in bilayer phosphorene under normal compressive strain}

\author{Aaditya Manjanath$^{\dagger,\ddagger,\P}$, Atanu Samanta$^{\dagger,\P}$,Tribhuwan Pandey$^{\dagger,\P}$ and Abhishek K. Singh$^{\dagger}$}
\address{$^{\dagger}$Materials Research Centre, Indian Institute of Science, Bangalore 560012, India}
\address{$^{\ddagger}$Centre for Nano Science and Engineering, Indian Institute of Science, Bangalore 560012, India}
\address{$^{\P}$These authors contributed equally to this work}
\ead{abhishek@mrc.iisc.ernet.in}

\begin{abstract}
Phosphorene, a two-dimensional (2D) analog of black phosphorous, has been a subject of immense interest recently, due to its high carrier mobilities and a tunable bandgap. So far, tunability has been predicted to be obtained with very high compressive/tensile in-plane strains, and vertical electric field, which are difficult to achieve experimentally. Here, we show using density functional theory based calculations the possibility of tuning electronic properties by applying normal compressive strain in bilayer phosphorene. A complete and fully reversible semiconductor to metal transition has been observed at $\sim13.35\%$ strain, which can be easily realized experimentally. Furthermore, a direct to indirect bandgap transition has also been observed at $\sim3\%$ strain, which is a signature of unique band-gap modulation pattern in this material. The absence of negative frequencies in phonon spectra as a function of strain demonstrates the structural integrity of the sheets at relatively higher strain range. The carrier mobilities and effective masses also do not change significantly as a function of strain, keeping the transport properties nearly unchanged. This inherent ease of tunability of electronic properties without affecting the excellent transport properties of phosphorene sheets is expected to pave way for further fundamental research leading to phosphorene-based multi-physics devices. 

\end{abstract}

\submitto{\NT}
\maketitle

\section{Introduction}
The experimental exfoliation of many-layer phosphorene~\cite{carrier_mobility_phos,phos_exp2,Liu-2014} has garnered immense interest, due to the presence of a direct bandgap~\cite{Liu-2014,GW_band_gap} and high carrier mobilities ($\sim1000$ cm$^2$V$^{-1}$s$^{-1}$)~\cite{carrier_mobility_phos}, making it a potential candidate for nanoelectronic devices. Within a phosphorene sheet, every phosphorous atom is covalently bonded with three neighboring atoms forming a highly corrugated honeycomb-like structure. These sheets are bonded by van der Waals (vdW) interactions in multi-layered phosphorene. In order to widen the range of applications of phosphorene, it is necessary to engineer the bandgap. It has been shown theoretically, that phosphorene exhibits strain-dependent electronic properties~\cite{D-I_transition_inplane}. The uniaxial in-plane strain leads to a direct to indirect bandgap (D-I) transition at very high tensile/compressive strains~\cite{D-I_transition_inplane}. In addition, applying normal compression~\cite{S-M_transition_monolayer} or biaxial strains~\cite{Cahir2014} on monolayer phosphorene also leads to modulation of its electronic structure.  

The bandgap tunability study has also been extended to bilayer and many-layer phosphorene. For example, the bandgap varies very strongly with the number of layers, with a range of $\sim2$ eV (monolayer) to $0.3$ eV (bulk)~\cite{Qiao2014,Zhang2014}. It has been recently reported that the bandgap in bilayer phosphorene can be modulated by changing the stacking order or by applying a vertical electric field~\cite{phos-stacking}. Experimentally, applying an electric field on such a system is difficult because the fields required are large ($\sim0.5$ V/\AA)~\cite{phos-stacking}. Another way of modulating the bandgap could be by application of experimentally feasible normal compressive (NC) strain, which is easiest to achieve due to relatively weaker vdW interlayer interactions as shown for other layered materials~\cite{Bhattacharyya2012,nayak2014,nanotech,atanu2014,pandey-2014}. Here, we show, using density functional calculations, the bandgap variation as a function of NC strain. Reversible direct to indirect and semiconductor to metal transitions, occur at applied NC strains of $3.32\%$ and $13.35\%$, respectively. In addition, we find significant anisotropy in the carrier mobilities and effective mass along $x$- and $y$-directions, which do not change much with the NC strain, thereby maintaining the excellent transport properties. Phonon spectra as a function of NC strain shows no instabilities, maintaining structural and mechanical integrity even at very high compressions. Furthermore, Raman frequency shift in A$_{\mathrm{1g}}$, B$_{\mathrm{2g}}$, and A$_{\mathrm{2g}}$ as a function of applied strain were mapped out, which will also help to monitor the bandgap modulation experimentally.

\begin{figure}[ht!]
\includegraphics[width=\columnwidth]{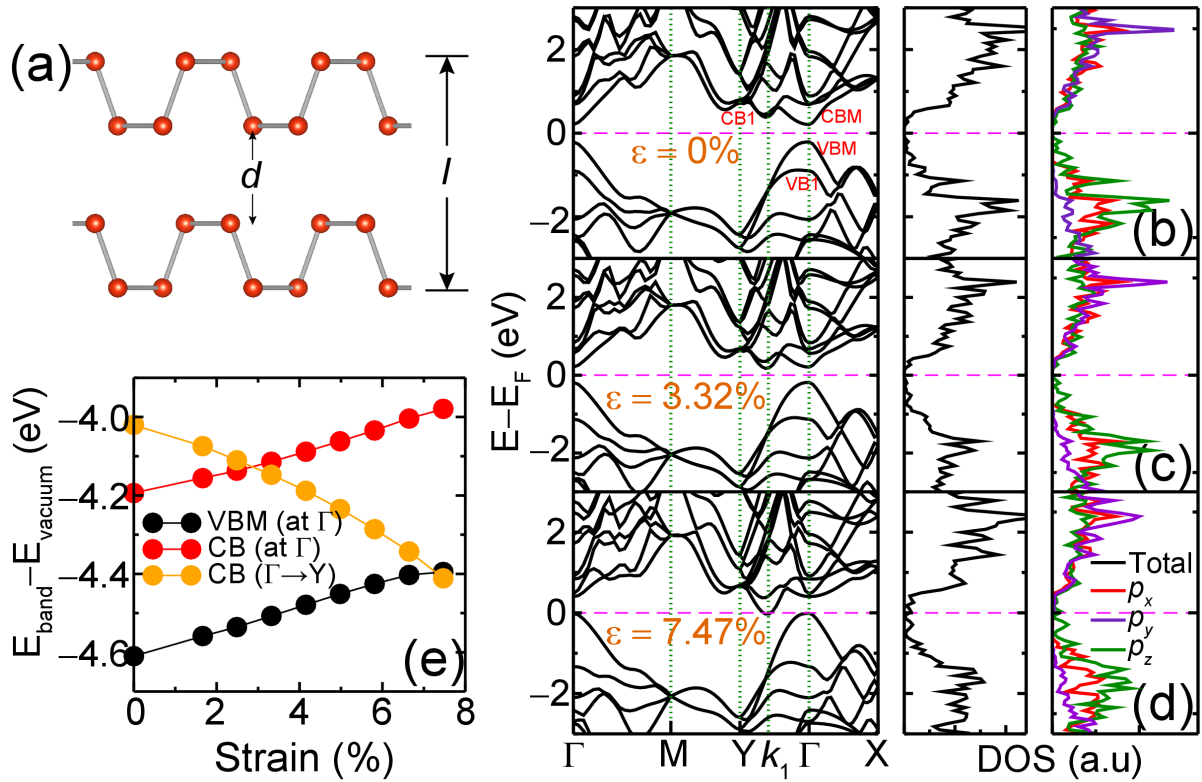} 
\caption{a) The crystal structure of bilayer phosphorene, $d$ and $l$ are the interlayer distances measured with respect to inner and outer atoms, respectively. Band structure and DOS of AB stacked phosphorene at b) $0\%$ c) $3.32\%$, and d) $7.47\%$ normal compression, are shown. e) The change in VBM and CBM with strain.} 
\label{fig:1} 
\end{figure}

\begin{figure}[!t]
\includegraphics[width=\textwidth]{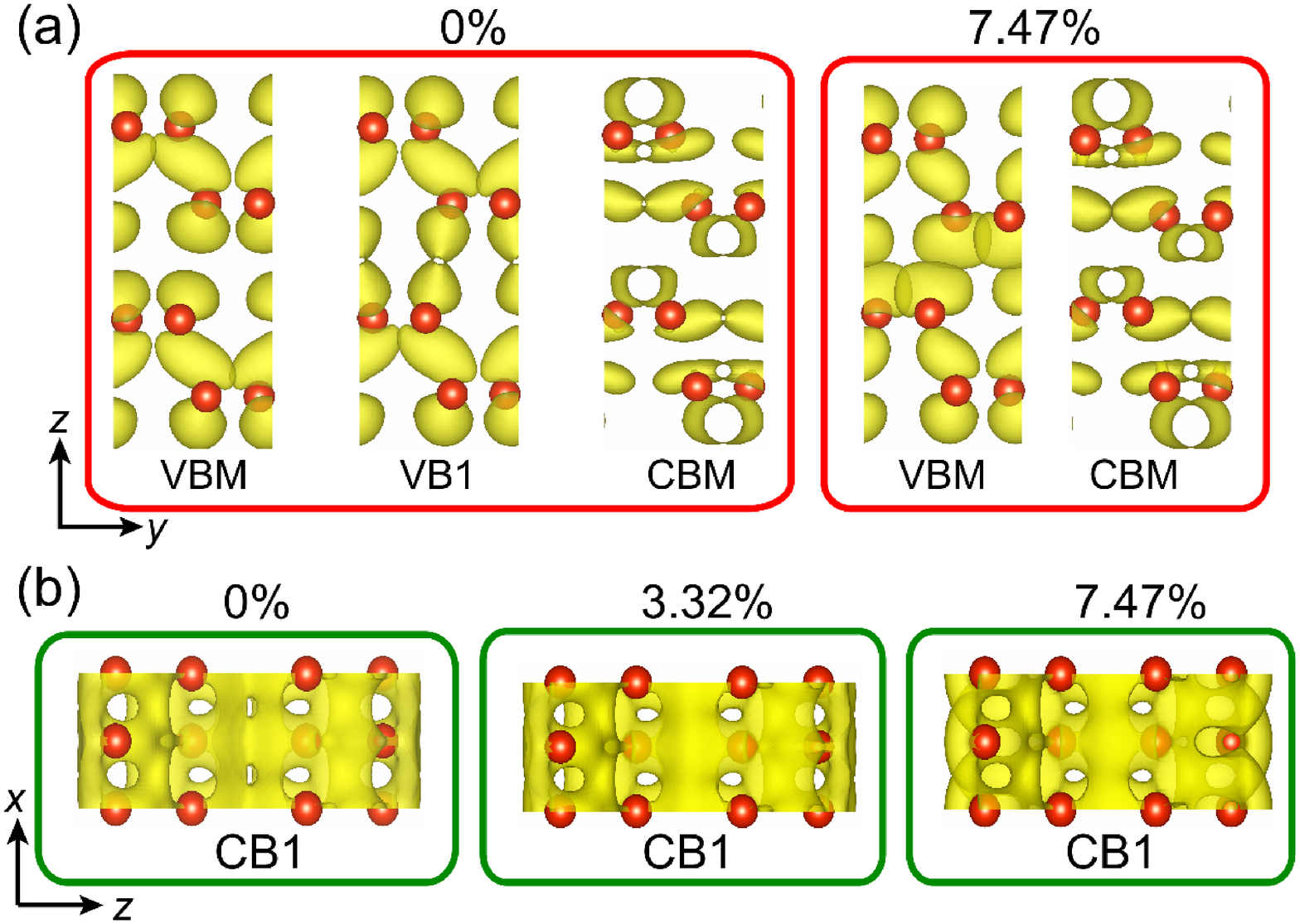} 
\caption{(a) Left panel shows band-decomposed charge densities corresponding to VBM, VB1, and CBM at $0\%$, and right panel, VBM and CBM at $7.47\%$ strain, as viewed in the $y$-$z$ plane. (b) The band-decomposed charge density of CB1 at $0$, $3.32$ and $7.47\%$ strains, respectively as viewed in the $x$-$z$ plane. All are plotted with the same isosurface value.} 
\label{fig:2} 
\end{figure} 
\section{Methodology}
The calculations were performed using first-principles density functional theory (DFT)~\cite{KohnSham} as implemented in the Vienna \textit{ab initio} simulation package (VASP)~\cite{Kresseetal,KresseetalPRB}. 
The layers in phosphorene are held together by vdW forces, which is incorporated by adding a semi-empirical dispersion potential ($D$) to the conventional Kohn-Sham DFT energy, through a
pair-wise force field following Grimme's DFT-D3 method~\cite{Grimme2010}. Projector augmented wave (PAW)~\cite{Blochl,KresseJoubert} pseudopotentials were used for the electron-ion interactions. The exchange and correlation part of the total energy was approximated by the generalized gradient approximation (GGA) using Perdew-Burke-Ernzerhof (PBE) type of functional~\cite{GGA-PBE}. Bilayer phosphorene has been predicted to exist in three different stacking, namely AA, AB, and AC. The AB-stacked structure is energetically the most preferred, and therefore, we carry out the rest of the study on this. The unit cell consists of eight atoms as shown in Fig.~\ref{fig:1}(a) and Fig. S1, with an optimum interlayer distance of $d=3.23$~\AA. A vacuum of $27$~\AA~is included along the $z$-axis to avoid spurious interactions between the periodic images. The Brillouin zone was sampled with a well-converged Gamma-centered \textbf{k}-mesh of $17\times15\times1$.

NC strain on this system was modelled by fixing the bottom layer and varying the $z$ co-ordinates of the atoms belonging to the top layer. A constrained structural relaxation was performed using the conjugate-gradient method until the absolute value of the components of forces were converged to within $0.005$ eV/\AA. Phonon dispersions and the Raman active modes were obtained from density functional perturbation theory (DFPT)~\cite{GonzeVigneron} with a strict energy convergence criterion of $1\times10^{-8}$ eV in order to obtain accurate forces. The Raman spectra were obtained by the linear response theory as implemented in Quantum {\scshape Espresso} package~\cite{QE}.
\section{Results and Discussion}
\begin{figure}[!ht]
\centering
\includegraphics[scale=1]{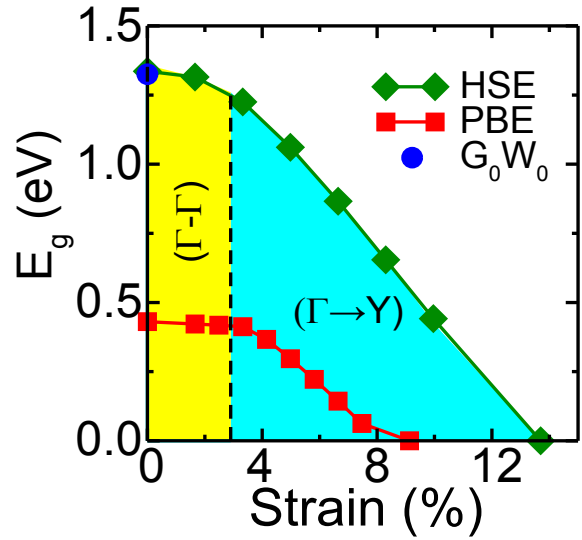} 
\caption{Bandgap change as a function of strain for PBE, HSE and G$_0$W$_0$ methods. The G$_0$W$_0$ bandgap at $0\%$ strain is from Ref.~\cite{Tran2014}}
\label{fig:3} 
\end{figure} 
Fig.~\ref{fig:1} (a) shows AB bilayer phosphorene crystal structure. The calculated optimized structural parameters using PBE including vdW interactions are shown in Table ST1, which are in good agreement with the previously reported results~\cite{Qiao2014}. The NC strain was calculated as $\epsilon=(l-l_{0})/l_{0}$, where $l_{0}$ and $l$ are the equilibrium and strained layered distance (Fig.~\ref{fig:1}(a)), respectively. The bond lengths $\delta_{1}$ and $\delta_{2}$ remain nearly unchanged over the entire range of strain values. The changes in the structural parameters under application of NC strain are shown in Fig. S1 (a), (b), and (c). There are no significant changes in $\alpha_1^\prime$, $\alpha_2$, and $\alpha_2^\prime$ observed with applied NC strain. The intralayer in-plane angle $\alpha_1$ (as shown in Fig. S1 (c)) shows relatively larger changes indicating that NC also enhances intralayer interactions. 

Next we study the effect of NC strain on the electronic properties of bilayer phosphorene. The PBE band structure calculations show that the unstrained AB stacked bilayer phosphorene has a direct bandgap at $\Gamma$ point of $0.48$ eV. In comparison to monolayer, the bandgap of bilayer phosphorene is 0.33 eV lower and shows two additional bands along with the original valence band (VB) and conduction band (CB). There are two valleys in the conduction band with a minima at $\Gamma$ and $k_1=(0.3158,0.0,0.0)$. The CBM is at $\Gamma$ point and the minima of CB1 (CBM$-1$) occurs at $k_1=(0.3158,0.0,0.0)$ (Fig.~\ref{fig:1}(b)). At $0\%$ strain, the CBM and VBM mainly originate from \textit{p$_{z}$} orbitals, whereas the VB1 (VBM$-1$), from \textit{p$_{z}$} and \textit{p$_{x}$} orbitals. The CB1 band at $k_1$ has major contributions from \textit{p$_{z}$} orbitals with small contributions from \textit{p$_{x}$} and \textit{p$_{y}$} orbitals as shown in Fig.~\ref{fig:1}(b). The bilayer phosphorene shows direct gap at $\Gamma$ point upto $3\%$ NC strain. With a further increase of NC strain, the bandgap changes from direct (at $\Gamma$) to indirect (along $\Gamma$-Y) with CB1 at $k_1$ becoming minima. At $7.47\%$  NC strain, the bilayer phosphorene becomes metallic. Fig.~\ref{fig:1}(e) clearly shows that the direct/indirect and finally S-M transition is the result of the competition of near-band-edge states at $\Gamma$ and at $k_1$ symmetric points. With an increase in strain, the energy of CB1 at wave vector $k_1$ reduces rapidly and becomes lower than the CBM at $\Gamma$. Most interestingly, the CBM and VBM at $\Gamma$ increase by equal amount keeping the bandgap nearly same with increasing NC strain (Fig.~\ref{fig:1}(e)). The shift in CB1 at $k_1$, with strain is much larger than that in the competing VBM and CBM at $\Gamma$ (Fig.~\ref{fig:1}(e)). Upon removing the NC strain, the structure completely recovers its $0\%$ geometry. The band structure also returns to its original form, thereby demonstrating complete reversibility in S-M transition without compromising on the structural integrity.

To understand the different trends of near-band-edge states at different symmetric $\textbf{k}$-points, the band-decomposed charge density as a function of NC strain is analyzed and shown in Fig.~\ref{fig:2}. The VBM shows an antibonding character, whereas VB1 shows bonding-like character originating from interlayer interactions (Fig.~\ref{fig:2}(a)). Therefore, VB1 gets pushed down to lower energy. The CBM shows antibonding character arising from interlayer interactions. Due to similar antibonding nature of VBM and CBM (interlayer interaction), energy of these states increase nearly equally with increasing NC strain. However, CB1 has bonding feature due to interlayer interactions, which mainly originates from \textit{p$_{z}$} orbitals. With increasing NCS the interlayer distance decrease, which leads to increase in the overlap of \textit{p$_{z}$} orbitals between two layers and lowering the energy of CB1. Due to this increase of interlayer interaction the CB1 becomes much more dispersive. At the S-M transition NC strain of $7.47\%$, the density of states at Fermi energy arises from phosphorous \textit{$p_{z}$} orbitals, as shown Fig.~\ref{fig:1}(d). This indicates that the strong interlayer interaction of \textit{$p_{z}$} orbitals leads to S-M transition.

It is well-known that DFT with standard exchange-correlation functionals such as the local-density approximation (LDA) or the generalized gradient approximation (GGA) underestimates the bandgap of semiconductors~\cite{Sham1983,Perdew1985}. The PBE bandgap of bilayer phosphorene is significantly lower than the experimentally reported photoluminescence (PL) and G$_{0}$W$_{0}$ values of $1.3$ eV ~\cite{Zhang2014} and $1.32$ eV ~\cite{Tran2014}, respectively. Therefore, it is very difficult to predict accurate strain dependence of electronic properties within PBE. In order to correct the PBE bandgaps, we used a hybrid Heyd-Scuseria-Ernzerhof (HSE06)~\cite{Heyd2005,Janesko2009,Ellis2011,Henderson2011} functional. The functional is separated into a long-range and a short-range part with 1/4 of the PBE exchange replaced by the Hartree-Fock exact exchange and contains the full PBE correlation energy. The HSE06 generally corrects the bandgap underestimation problem by partially removing the self interaction. Previous studies have shown that the hybrid-functional HSE06 method shows significant improvement of bandgap of mono- and multi-layer phosphorene~\cite{Qiao2014,Tran2014,Peng2014}. Our HSE06 calculations at $0\%$ strain with a mixing parameter of 0.35 show a bandgap of 1.33 eV, which is in excellent agreement with G$_0$W$_0$ value~\cite{Tran2014} (Fig.~\ref{fig:3}). The dispersion of the valence band does not change with respect to PBE. However, the conduction band shifts upwards without significant change in the band dispersion (Fig. S2). Under NC strain, similar changes in bandgap are observed as shown in Fig.~\ref{fig:3}. However, due to a larger bandgap, S-M transition occurs at a higher strain value of $13.35\%$. We calculate the equivalent pressure value in GPa, using the formula $P=(E-E_0)/[(l-l_0)A]$, where $E$ and $E_0$ are the total energies of strained and unstrained structures, and $A$, the area of the sheet. The S-M strain of $13.35\%$ corresponds to $\sim8.9$ GPa, which can possibly be realized experimentally by using the diamond anvil cell technique, recently applied on multilayer MoS$_2$~\cite{nayak2014}. The mechanism leading to the S-M transition for HSE calculations is similar as the PBE calculations (Fig. S2). In addition, similar to PBE, the S-M transition in this case is completely reversible. Interestingly, even at $0\%$ strain, the interlayer interaction reduces the bandgap ($\approx$ 0.67 eV) compared to monolayer. This bandgap reduction is much higher than the other vdW-layer materials~\cite{Bhattacharyya2012}. Thus, interlayer interaction is an important and dominating parameter, which can be modified by applying the NC strain to tune the bandgap of bilayer phosphorene.

\begin{figure}[ht!]
\includegraphics[width=\columnwidth]{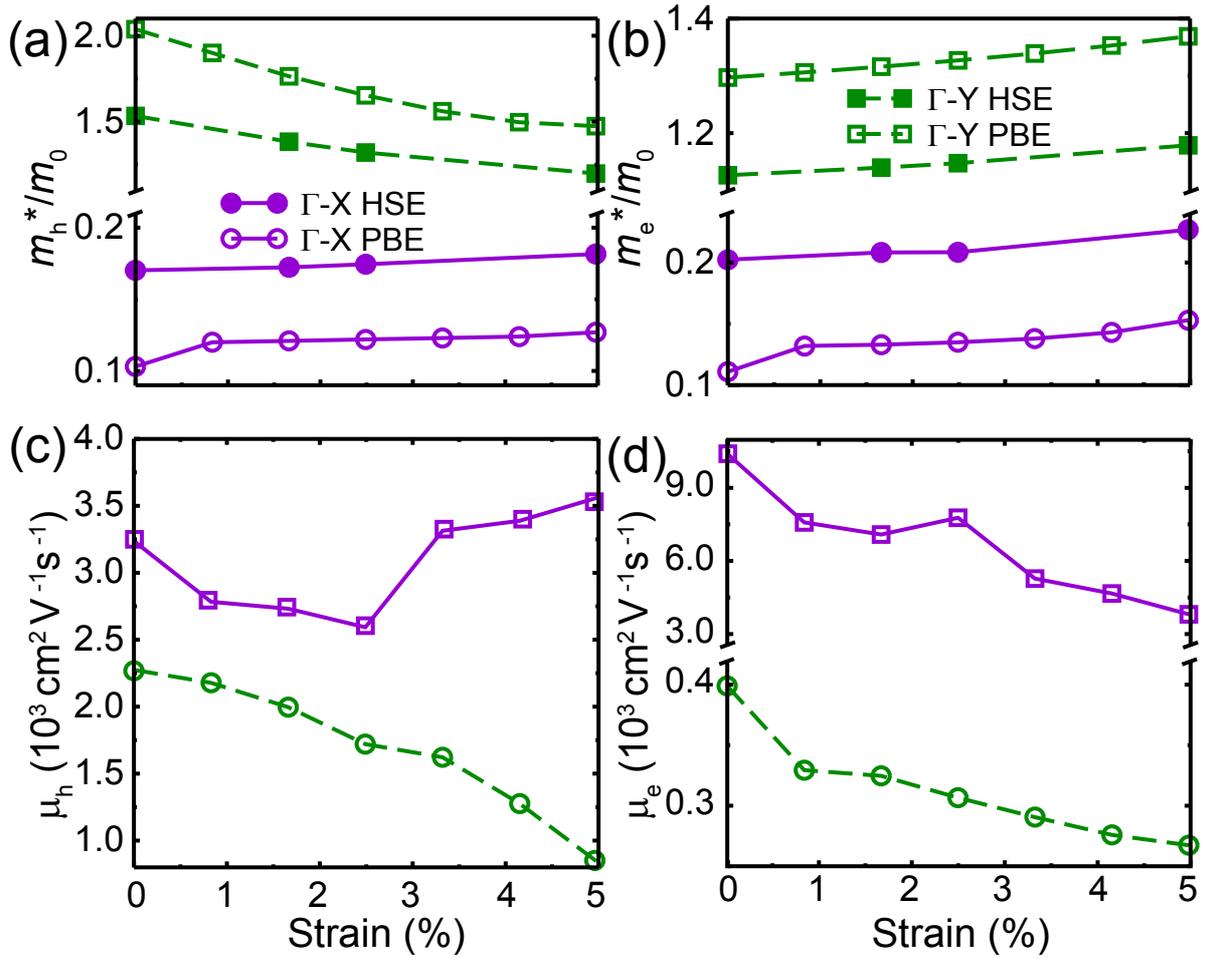} 
\caption{(a) and (b) Effective masses, (c) and (d) room temperature mobilities of holes and electrons, respectively as a function of NC strain along $\Gamma$-X and $\Gamma$-Y directions.} 
\label{fig:4} 
\end{figure}

Since phosphorene exhibits excellent transport properties, it is important to study the effect of NC strain on carrier effective masses and mobilities as the electronic properties are governed by these. We assume that the mechanism limiting the mobility of charge carriers is the scattering due to only longitudinal acoustic phonons~\cite{mob_calc_1,mob_calc_2}. The carrier mobility in a 2D system is given by~\cite{Qiao2014}:
\begin{equation}\label{mobility}
\mu_{\alpha}=\frac{e\hbar^3C_{s\alpha}}{k_BTm_{\alpha}^*\sqrt{m_{\alpha}^*m_{\beta}^*}E_{1\alpha}^2}
\end{equation}
where, $C_{s\alpha}$, $m_{\alpha}^*$, $T$ and $E_{1\alpha}$ are the in-plane stiffness, effective mass of the charge carrier (hole/electron), temperature and the deformation potential along the $\alpha^{\mathrm{th}}$ direction. $C_{s\alpha}$ is defined as:
\begin{equation}
C_{s\alpha}=\frac{1}{A_0}\frac{\partial^2E}{\partial\epsilon_\alpha^2}
\end{equation}
where, a uniaxial in-plane strain $\epsilon_\alpha=\frac{a_\alpha-a_{0\alpha}}{a_{0\alpha}}$; ($a_\alpha$ - lattice parameter under strain, and $a_{0\alpha}$ - equilibrium lattice parameter), is applied along the transport direction $\alpha$ ($\alpha=x$, $y$), and second-order derivative of energy is obtained by fitting the energy-strain relation to a second-order polynomial of the form:
\begin{equation}\label{quad-poly}
E=E_0+\tilde{C_1}\epsilon+\tilde{C_2}\epsilon^2
\end{equation}
The effective mass of a charge carrier is the inverse of the curvature of the band, i.e. $m_\alpha^*=\frac{\hbar^2}{\partial^2E/\partial k_\alpha^2}$. We calculate the effective masses in the neighborhood of the $\Gamma$ point along the X and Y directions. The valence band maximum and the conduction band at $\Gamma$ are, essentially, parabolic in this small interval, which can be fitted to a second-order polynomial of a similar form as eqn.~\ref{quad-poly}. The coefficient of the second-order derivative gives the curvature, which is utilized to compute the effective mass.
The deformation potential is the change in the potential as a consequence of the applied in-plane strain, which is defined as: $E_{1\alpha}=\frac{\Delta V_i}{\Delta a_\alpha/a_{0\alpha}}$, where, $\Delta V_i$ is the change in the energy of the $i^\mathrm{th}$ band under in-plane strain. All the parameters have been obtained within the PBE framework.

The parameters that markedly influence the mobility are the deformation potential $E_1$, the effective masses and the in-plane stiffness $C_s$. We analyze the effect of normal compression on each of these parameters in addition to the mobility. Figs.~\ref{fig:4}(a) and (b) show the effective masses of holes and electrons along the $\Gamma$-X (violet solid lines in Fig.~\ref{fig:4}) and $\Gamma$-Y (green dashed lines in Fig.~\ref{fig:4}) directions, as a function of normal compression. The effective masses have been calculated using both PBE (unfilled symbols in Fig.~\ref{fig:4}) and HSE06 functionals (filled symbols in Fig.~\ref{fig:4}) to understand the effect of bandgap underestimation. The PBE values of the effective masses are in good agreement in terms of the order of magnitude with the HSE06 ones. In addition, the trends of the effective masses are the same for both PBE and HSE06. Hence, qualitatively and to some extent, quantitatively, PBE is sufficient to analyze the transport properties. For both holes (Fig.~\ref{fig:4}(a)) and electrons (Fig.~\ref{fig:4}(b)), the effective masses along the $\Gamma$-Y direction are significantly higher than those along $\Gamma$-X, implying a high anisotropy in the system. On application of NC strain, the effective masses along $\Gamma$-Y ($\Gamma$-X) decrease (increase). The reason for this increase/decrease can be attributed to the decrease/increase in the curvature of the bands caused by the decrease/increase in the overlap of the bonding (holes)/antibonding (electrons) orbitals.

The in-plane stiffness, $C_{sx}$ and $C_{sy}$ along the $x$- and $y$-directions, respectively, were calculated according to the methodology described in our previous work~\cite{in-plane-stiff}. Although the values do not vary significantly with NC strain, overall, the $C_{sx}<C_{sy}$ (Table S2), implying anisotropy in the elastic properties. The trend is similar to that observed in monolayer phosphorene~\cite{D-I_transition_inplane}. Fig. S3(a) shows, the variation in total energy with respect to in-plane strain for a given normal compression. The high curvature in the energy surface for $\epsilon_y$, suggests that it is difficult to apply strain along the $y$-direction compared to $x$, also indicating that $C_{sy}$ is higher compared to $C_{sx}$. In contrast to this, the trends are different in the case of the deformation potentials (Fig. S4) depending on the charge carrier under consideration (Table S2). The values as a function of normal compression and the band-decomposed charge density plots for a given compression are included in Figs. S3(b)-(e). With respect to holes, the VB wavefunction along $y$ is slightly more localized compared to $x$-direction. Therefore, any structural deformations due to longitudinal phonons along this direction will affect the VB wavefunction lesser compared to $x$-direction, thereby leading to $E_{1y}$ being lower than $E_{1x}$. When comparing the deformation potentials of holes and electrons along a particular direction, we find contrasting trends, i.e. $E_{1x}^{\mathrm{holes}} > E_{1x}^{\mathrm{electrons}}$ and $E_{1y}^{\mathrm{holes}} < E_{1y}^{\mathrm{electrons}}$, which are once again, well explained by the band-decomposed charge density plots in Figs. S3(b)-(e). By utilizing the trends described into equation~\ref{mobility}, we find that for a given charge carrier, $\mu_x>\mu_y$, whereas, along a particular direction, the relationships are $\mu_x^{\mathrm{electrons}}>\mu_x^{\mathrm{holes}}$, $\mu_y^{\mathrm{electrons}}<\mu_y^{\mathrm{holes}}$. Hence, the overall trends in mobilities (Table S2) are controlled by the trends in effective masses and deformation potentials. The mobilities do not change significantly as a function of NC strain. Therefore, the transport properties of bilayer phosphorene are preserved, despite the application of NC strain.

\begin{figure}[ht!]
\includegraphics[width=\columnwidth]{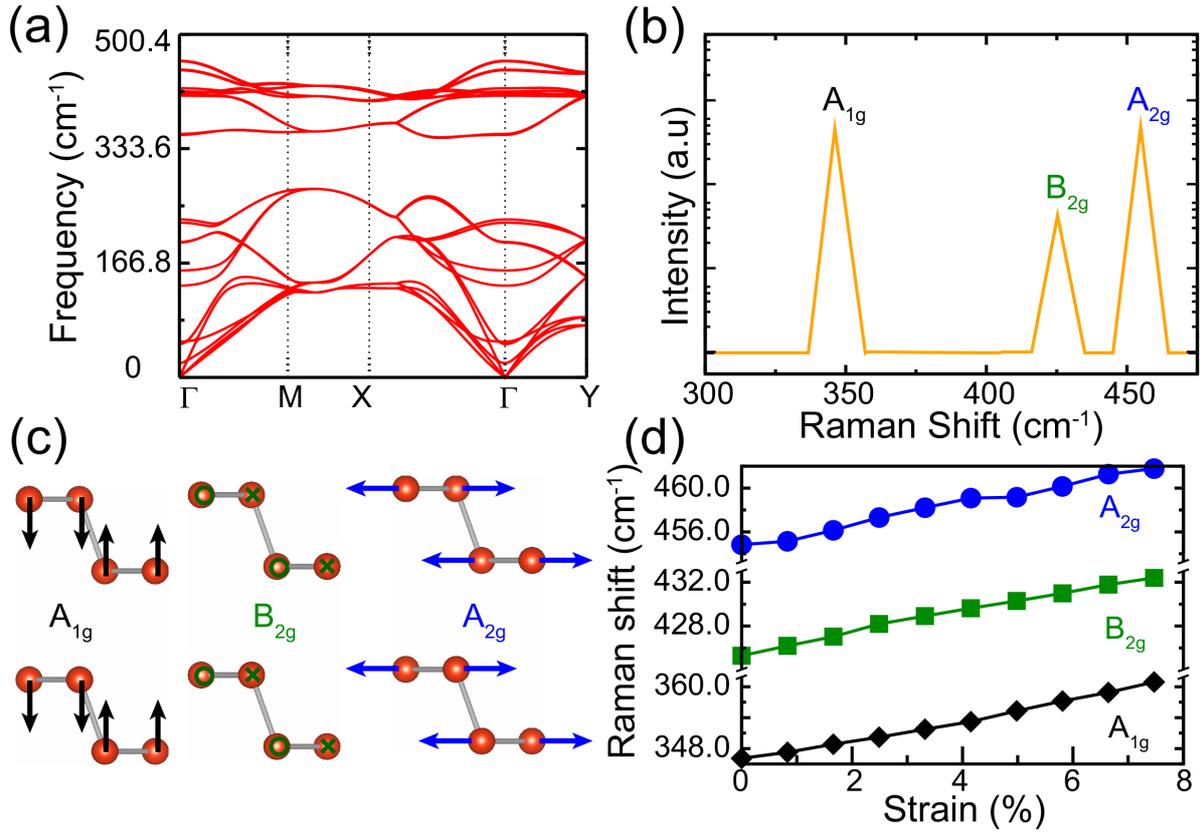} 
\caption{(a) Phonon Dispersions for unstrained bilayer phosphorene. (b) Simulated Raman spectra of unstrained phosphorene. For bilayer phosphorene, only three modes, A$_{1\mathrm{g}}$, B$_{2\mathrm{g}}$, and A$_{2\mathrm{g}}$, exhibit prominent Raman scattering peaks. (c) Representative vibrations involved in A$_{1\mathrm{g}}$, B$_{2\mathrm{g}}$, and A$_{2\mathrm{g}}$ modes. (d) Raman frequencies as a function of NC strain for the three Raman active modes.} 
\label{fig:5} 
\end{figure}

To explore the structural integrity of bilayer phosphorene under NC strain, phonon calculations were performed. The symmetry of bilayer phosphorene is described by the D$_{2\mathrm{h}}$ point group. Fig.~\ref{fig:5}(a) shows the calculated phonon dispersion of bilayer-phosphorene, which has 24 phonon branches. The optical branches have small splitting and are nearly double degenerate due to the weak interlayer interaction. From Fig.~\ref{fig:5}(a), we can see that there is no imaginary frequency in the full phonon spectra, indicating the dynamical stability of bilayer phosphorene. This conclusion is in good agreement with experiments, where it is shown that the exfoliated flakes of black phosphorous are stable even in free-standing form~\cite{phos-2d}. The phonon dispersion calculation as a function of strain indicates the dynamical stability of bilayer phosphorene even at high strains (Fig. S5). At a very high strain of $\sim15\%$ small instabilities appear in the phonon dispersion, which is well above the critical strain (the strain at which S-M transition occurs) predicted by PBE or HSE06. Based on this analysis we conclude that the electronic properties of bilayer phosphorene can be tuned efficiently and reversibly under a large strain window while maintaining structural integrity. 

The key changes in lattice vibrational modes as a function of NC strain can be captured by Raman spectra~\cite{nayak2014}. We calculate the Raman spectra as a function of NC strain as shown in Fig.~\ref{fig:5}(b) and Fig. S6. Here, only three modes, A$_{1\mathrm{g}}$, B$_{2\mathrm{g}}$, and A$_{2\mathrm{g}}$, exhibit prominent Raman scattering peaks. The out-of-plane A$_{1\mathrm{g}}$ mode occurs due to opposing vibrations of top and bottom P atoms with respect to each other within the same layer. The B$_{2\mathrm{g}}$ and A$_{2\mathrm{g}}$ mode is associated with the in-plane vibration of the P atoms in directions opposite to one another as shown in Fig.~\ref{fig:5}(c). As shown in Fig.~\ref{fig:5}(b) these peaks are located at $346.95$, $425.62$, and $454.63$ cm$^{-1}$ for unstrained phosphorene, which are in good agreement with the reported experimental values~\cite{phos-nanores,Liu-2014}. The strain dependence of the three most prominent Raman active modes (one out-of-plane (A$_{1\mathrm{g}}$) and two in-plane (B$_{2\mathrm{g}}$ and A$_{2\mathrm{g}}$)) is shown in Fig.~\ref{fig:5}(d). While all the three different modes show qualitatively, the same behavior with respect to applied strain and exhibit a linear increase (blue shift), the rate of increase is different for different modes. The A$_{1\mathrm{g}}$ mode increases at a rate of 2.04 cm$^{-1}$/$\Delta\epsilon$ and exhibits the highest blue shift of $\sim 15$cm$^{-1}$ under the strain range investigated here. The other two modes B$_{2\mathrm{g}}$ and A$_{2\mathrm{g}}$ also increase linearly but with a lower rate of $0.95$ and $0.96$ cm$^{-1}$/$\Delta\epsilon$, respectively. The different rate of increase of Raman active modes can be explained by analyzing the type of vibrations involved in these modes. The A$_{1\mathrm{g}}$ mode emerges from out-of-plane vibrations of phosphorus atoms and has a strong dependence on interlayer distance. The NC strain reduces the interlayer distance leading to a strong blue shift in the A$_{1\mathrm{g}}$ mode, whereas the B$_{2\mathrm{g}}$ and A$_{2\mathrm{g}}$ modes arise from the in-plane vibrations of phosphorus atoms and hence experience lower shifts. Therefore, the Raman spectra and its trend with respect to NC strain can be an effective way of monitoring the changes in the electronic structure and lattice virations in bilayer phosphorene, experimentally.

\section{Conclusion}
In conclusion, we report reversible semiconductor to metal transition in bilayer phosphorene by applying NC strain. The bandgap can be reversibly tuned within a large energy range from 1.32 eV to 0 eV. We demonstrate that strong interlayer interactions between phosphorus-$p_{z}$ orbitals causes the S-M transition under NC strain. An estimate of the effect of NC strain on transport properties through the computation of carrier effective mass and mobility shows that the transport properties do not change significantly as a function of strain. Therefore, the transport properties remain nearly unchanged. Furthermore, we calculate the Raman spectra as a function of strain and identify the prominent Raman active modes (A$_{1\mathrm{g}}$, B$_{2\mathrm{g}}$ and A$_{2\mathrm{g}}$) which increase linearly with NC strain. Due to strong interlayer interaction the rate of increase was found to be highest for the out-of-plane A$_{1\mathrm{g}}$ mode. Raman spectra can serve to monitor the changes in the electronic structure and vibrational properties, experimentally. This reversible bandgap tuning by application of NC strain opens up a window for phosphorene to a wide range of applications such as pressure sensors, optoelectronic and many other nanoscale devices.

\section*{Supporting Information}
Structural details and transport properties are tabulated in the Supplementary Tables S1 and S2. The changes in the geometry, HSE06 electronic structures as a function of NC strain, the energy vs. strain, the characteristics of the bands, electron and hole deformation potentials, phonon dispersion as a function of NC strain, and Raman shift as a function of NC strain, are included here. This material is available free of charge.

\section*{Acknowledgements}
The authors acknowledge financial support from DST Nanomission. The authors thank the Materials Research Centre and Supercomputer Education Research Centre for the required computational facilities.

%\section*{References}
%\bibliographystyle{iopart-num}
%\bibliography{paper-nano}
%\section*{References}
%\providecommand{\newblock}{}
%\begin{thebibliography}{10}
%\expandafter\ifx\csname url\endcsname\relax
%  \def\url#1{{\tt #1}}\fi
%\expandafter\ifx\csname urlprefix\endcsname\relax\def\urlprefix{URL }\fi
%\providecommand{\eprint}[2][]{\url{#2}}
\section*{References}
%\bibliographystyle{iopart-num}
%\bibliography{manuscript}
\providecommand{\newblock}{}

\end{document}